\documentclass[cameraready]{Interspeech}

\title{Towards Array-Invariant Speech Enhancement \\ via Geometry-Aware Dynamic Convolution}

\author[affiliation={1},orcid=0009-0006-3717-4444]{Zhenglong}{Liu}
\author[affiliation={1,2},orcid=0000-0003-4500-3515]{Wangyou}{Zhang}
\author[affiliation={1,2},orcid=0000-0003-0299-9914]{Chenda}{Li}
\author[affiliation={1,2},orcid=0000-0002-0314-3790]{Yanmin}{Qian}

\address{
    $^1$ Auditory Cognition and Computational Acoustics Lab\\
    Shanghai Jiao Tong University, Shanghai, China\\
    $^2$ VUI Labs
}

\email{zhenglong.liu@sjtu.edu.cn, wyz-97@sjtu.edu.cn, lichenda1996@sjtu.edu.cn, yanminqian@sjtu.edu.cn}

\keywords{Multi-channel speech enhancement, microphone array invariant, dynamic convolution}

\usepackage{comment}
\usepackage{multirow}
\usepackage[table]{xcolor}

\begin{document}

\maketitle

\begin{abstract}
Multi-channel speech enhancement (SE) systems exhibit superior performance over single-channel methods but are constrained to fixed microphone array configurations. This restricts their real-world deployment across devices with diverse array geometries. While recent array-agnostic SE methods address variable microphone numbers and permutations, they largely fail to exploit explicit array geometry priors when available, missing a crucial cue for optimal spatial filtering. A Geometry-Aware Dynamic Convolution (Geo-DConv) framework is proposed, which explicitly leverages microphone coordinates to transform standard fixed-array SE models into robust array-invariant systems. Experiments are conducted on the recent real-recorded RealMAN multi-channel speech dataset. Results demonstrate that the proposed architecture enables two widely used fixed-array models to adapt to array-invariant settings, with consistent performance improvements across diverse array topologies.
\end{abstract}

\section{Introduction}
While multi-channel speech enhancement (SE) theoretically offers a higher performance upper bound than single-channel approaches, its real-world applicability is often hindered by its reliance on fixed microphone array geometries. The geometric variations across different arrays make it difficult to merge existing datasets into a unified, large-scale training corpus. As a result, models frequently require device-specific retraining. This contrasts sharply with single-channel SE models, which can achieve robust generalization across various distortions \cite{li2025sense,Zhang2025AnyEnhance} by scaling up the training data. 

Array-agnostic SE has been proposed to overcome this limitation. These approaches must address two primary challenges: the variable microphone numbers and their arbitrary permutations. Existing solutions generally fall into two categories. In the first category, multi-channel inputs are processed through batch operations, then extracted cross-channel relationships by dimension-independent mechanisms, and merged into a single-channel output by setting a reference channel or taking the average. Representative methods are Transform-Average-Concatenate (TAC) \cite{luo_end--end_2020}, self-attention-based models \cite{Pandey2022TPARN}, and Transform-Attention-Concatenate (TA\textsubscript{tt}C) \cite{zhang_improving_2024}.
The second category transforms arbitrary array inputs into fixed-dimensional multi-channel signals before processing. For instance, converting recordings into First-Order Ambisonics (FOA) \cite{tatarjitzky_ambidrop_2025} results in a fixed 4-channel representation, regardless of the number of microphones employed. However, this method cannot be utilized for arrays consisting of fewer than four microphones, as underdetermined conversion equations are produced. Recently, UniArray \cite{chen_uniarray_2025} introduced virtual microphone estimation to map an arbitrary number of signals to a fixed dimensionality via interpolation-based upscaling. While this design ensures consistent input dimensions for the subsequent neural network, it still requires input order permutation as an essential preprocessing step.

Despite their flexibility, array-agnostic SE models generally perform worse than fixed-array models for two main reasons. Firstly, in fixed-array methods, cross-channel feature extraction typically begins from the very first layer, and many effective modules have been designed to model these relationships. In contrast, array-agnostic SE models initially extract time-frequency features for each channel separately through batch operations, which leads to a lack of methodological diversity in modeling cross-channel relations. Secondly, during the training of fixed-array models, all recordings share the same array geometry, allowing the model to learn a favorable spatial bias corresponding to the specific device. However, existing array-agnostic methods fail to exploit explicit array geometry information, relying mostly on generic time-frequency features and implicit inter-channel correlations.

The immense value of explicit geometric information, however, has been well-established across various speech processing domains. In traditional signal processing, the Minimum Variance Distortionless Response (MVDR) \cite{Darzi2016MVDR} algorithm explicitly utilizes microphone array spacing to formulate the steering vector, demonstrating the fundamental benefit of geometry for SE. More recently, in the deep learning domain, array geometry cues have been successfully leveraged in geometry-invariant Direction-of-Arrival (DOA) estimation. For instance, GI-DOAEnet \cite{baek_dnn-based_2025} employs microphone positional encodings (MPEs) to inject unique spatial information by modulating microphone spherical coordinates via sinusoidal functions. Despite these proven benefits, explicit array geometry cues remain largely unexploited in the design of array-agnostic SE models. Furthermore, since recording devices are commonly specified in multi-channel speech datasets, acquiring such geometric configurations requires absolutely no additional annotation effort, making it a highly feasible yet neglected direction for SE improvement.

In this work, we explicitly leverage the array geometry as additional input of the network through the proposed geometry-aware dynamic convolution block, which enables arbitrary fixed-array SE models to be converted to array-invariant models. Furthermore, by training on real-recorded array datasets, we avoid the common issue of mismatch between simulated data and real-world environments in array-based speech enhancement.
\section{Methods}
\begin{figure}[t] 
    \centering
    \includegraphics[width=0.99\linewidth]{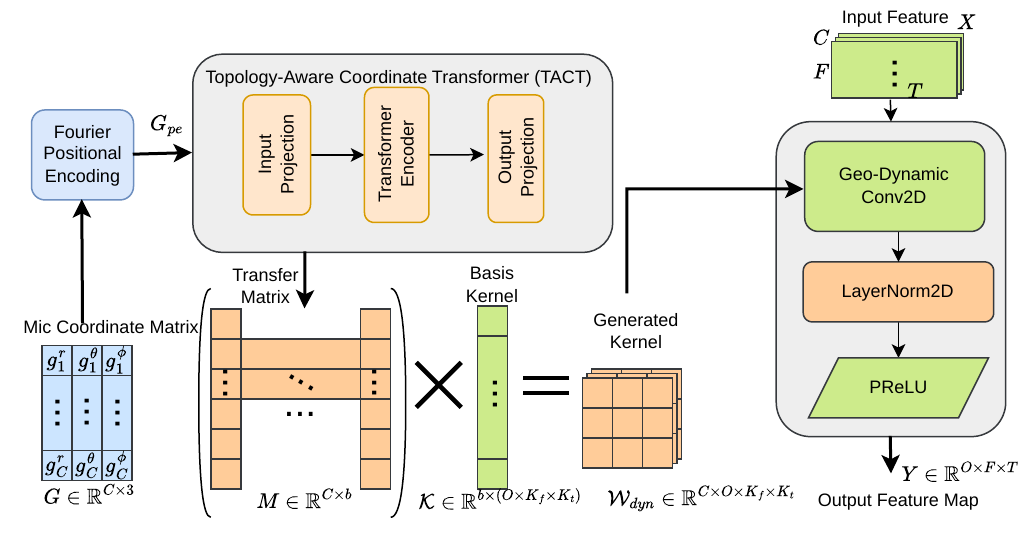} 
    \caption{Architecture of the proposed Geo-DConv. The left part illustrates the Geometry-Aware Dynamic Kernel Generation based on the Topology-Aware Coordinate Transformer (TACT). The right part shows the Dynamic Convolution process applied to input acoustic features.}
    \label{fig:model_arch}
\end{figure}
In common fixed-array SE, cross-channel modeling typically begins at the very first convolutional layer. Input features are stacked along the channel dimension, and inter-channel information is fused within the layer to form the output~\cite{Beam_TasNet-Ochiai2020,Closing-Zhang2021}. However, a fundamental limitation of conventional convolutional layer is its strict requirement for a fixed input dimension. This inherent rigidity prevents the direct application of standard convolutions in array-invariant SE systems. To bridge this gap, a dynamic convolution mechanism is designed to accommodate arbitrary input channel dimensions and spatial permutations. 

The overall architecture of the proposed method is depicted in Figure~\ref{fig:model_arch}. First, the relative coordinates of the arbitrary microphone array are transformed using Fourier positional encoding. These encoded representations are subsequently fed into a novel Topology-Aware Coordinate Transformer (TACT) module, which yields a dynamic transformation matrix. By applying matrix multiplication, this matrix dynamically adapts the weights of the fixed-dimension convolution kernels to align with the specific input configuration of the target array. Following this geometry-guided adaptation, the features are processed through Layer Normalization (LN) and an activation function before being propagated to the subsequent neural network layers. By leveraging explicit array geometry, this mechanism serves as a universal adapter, effectively transforming conventional fixed-array algorithms into array-invariant models.
\subsection{Problem Formulation}
Without loss of generality, the enhancement is assumed to operate in the time-frequency domain. Let $X \in \mathbb{R}^{C \times F \times T}$ denote the multi-channel acoustic feature (the real and imaginary parts of STFT complex spectrograms are concatenated along the frequency dimension), where $C$ represents the variable number of microphones, and $F, T$ represent the frequency and time dimensions, respectively. Each microphone is associated with a relative coordinates $g_i \in \mathbb{R}^3$, forming a coordinate matrix $G = [g_1, g_2, \dots, g_C]^\top \in \mathbb{R}^{C \times 3}$. The coordinates can be formatted in either Cartesian $(x, y, z)$ or Spherical $(r, \theta, \phi)$ systems. A Geometry-Aware Dynamic Convolution (Geo-DConv) layer is designed to handle variable-channel inputs and incorporate array geometry information, as formulated below,
\begin{equation}
\label{eq:dynconv}
Out=\text{Geo-DyncConv}(G,X)
\end{equation}

\subsection{Geometry-Aware Dynamic Convolution (Geo-DConv)}
To process an arbitrary number of microphones without structural constraints, Geo-DConv is proposed. A typical convolution kernel is defined as $\mathcal{K} \in \mathbb{R}^{b \times O \times K_f \times K_t}$, where $b$ denotes the basis input dimension, $O$ is the fixed output channel size, and $K_f, K_t$ are the kernel size in frequency and time dimensions.

The dynamic convolutional weight $\mathcal{W}_{dyn}$ for a specific array geometry is generated through a linear combination of the basis kernels $\mathcal{K}$, guided by a dynamic transformation coefficient matrix $M \in \mathbb{R}^{C \times b}$. Specifically, the combined kernel is formulated as:
\begin{equation}
\label{eq:dynamic_weight}
\mathcal{W}_{dyn}^{(c, o, :, :)} = \sum_{j=1}^{b} M_{c, j} \cdot \mathcal{K}^{(j, o, :, :)}
\end{equation}
where $\mathcal{W}_{dyn} \in \mathbb{R}^{C \times O \times K_f \times K_t}$. The core challenge therefore lies in designing a robust mechanism to generate the matrix $M$ from the relative coordinates of the array $G$, which must capture complex spatial interactions while maintaining adaptability to variable array configurations.

\subsection{Topology-Aware Coordinate Transformer (TACT)}
Inspired by Implicit Neural Representations (INR) in NeRF~\cite{mildenhall2020nerf} and NAF~\cite{luo_NAF_2022}, Fourier Positional Encoding (PE) is introduced to better characterize fine-grained variations in coordinates:
\begin{equation}
\label{eq:fourier_pe}
\begin{split}
\gamma(g_i) = \big[\, 
& g_i, \sin(2^0 \pi g_i), \cos(2^0 \pi g_i), \dots, \\
& \sin(2^{L-1} \pi g_i), \cos(2^{L-1} \pi g_i) 
\,\big]
\end{split}
\end{equation}
where $L$ is the number of frequency bands, yielding an encoded matrix $G_{pe} \in \mathbb{R}^{C \times d_{pe}}$ ($d_{pe} = 3 + 6L$).

The Topology-Aware Coordinate Transformer (TACT) is proposed to perform global array topology modeling. The encoded coordinate matrix $G_{pe}$ is first projected into the hidden dimension $d_{\text{hidden}}$:
\begin{equation}
\label{eq:g0}
G^{(0)}=G_{pe} W_{in}, \quad W_{in} \in \mathbb{R}^{d_{pe} \times d_{\text{hidden}}}
\end{equation}
The projected matrix $G^{(0)}$ is treated as a sequence of $C$ tokens and fed into a Transformer Encoder block. The topology-aware feature representation $Z$ is computed via the Multi-Head Self-Attention (MHSA) mechanism:
\begin{equation}
\label{eq:tact_qkv}
Q = G^{(l)}W_Q, \quad K = G^{(l)}W_K, \quad V = G^{(l)}W_V
\end{equation}
\begin{equation}
\label{eq:tact_mhsa}
Z^{(l+1)} = \text{LayerNorm}\left(Z^{(l)} + \text{MHSA}(Q, K, V)\right)
\end{equation}
Subsequently, the final transformation coefficient matrix $M$ is obtained through a linear output projection after $L_{layers}$ encoding layers:
\begin{equation}
\label{eq:tact_output}
M = Z^{(L_{layers})} W_{out}, \quad W_{out} \in \mathbb{R}^{d_{hidden} \times b}
\end{equation}

\textbf{Permutation Equivariance and Stability:} 
A critical property of the TACT module is its ability to maintain output stability under varying input channel permutations. Let $P \in \{0,1\}^{C \times C}$ be a permutation matrix representing a specific spatial ordering of the microphones. If the input coordinates are permuted as $PG$, the point-wise nature of the Fourier PE and the permutation-equivariant property of the MHSA mechanism ensure that the generated transformation matrix becomes $PM$. According to Eq.~\ref{eq:dynamic_weight}, the resulting dynamic convolution kernel is correspondingly permuted to $P\mathcal{W}_{dyn}$ along its input channel dimension. When performing the convolution operation, the dot product between the permuted input features $PX$ and the permuted weights $P\mathcal{W}_{dyn}$ remains mathematically invariant ($PX \circledast P\mathcal{W}_{dyn} = X \circledast \mathcal{W}_{dyn}$). This inherently guarantees that the extracted feature representations remain stable and consistent, regardless of the random ordering of input channels in practical scenarios.

\subsection{Overall Architecture and Integration}
\label{ssec:overall_architecture}
For compatibility with downstream fixed-array algorithms, LN and PReLU activation are adopted, with the overall architecture formulated as follows.
\begin{equation}
\label{eq:overall_block}
Y = \text{PReLU}\left( \text{LayerNorm}\left( \text{Geo-DyncConv}(G,X) \right) \right)
\end{equation}
This design maps variable-dimensional inputs to fixed-dimensional outputs, which are then processed by subsequent fixed-array algorithms.

\section{Experiments}

\subsection{Datasets}

\begin{table*}[!t]
\centering
\caption{Performance and efficiency comparison on the RealMAN dataset. Models are grouped by their training and application scenarios. In the \textit{Geometry-Invariant} setting (our main focus), the \textbf{best} results are highlighted in bold, and the \underline{second-best} are underlined. Our proposed methods are denoted in \textbf{bold} with `(Ours)'. Fixed-array methods serve as a geometry-specific upper bound.}
\label{tab:main_results}
\small
\setlength{\tabcolsep}{4pt}
\begin{tabular}{c l cc cccccccc}
\toprule
\multirow{2}{*}{\textbf{No.}} & \multirow{2}{*}{\textbf{Model}} & \multirow{2}{*}{\shortstack{\textbf{\#Params}\\(M)}} & \multirow{2}{*}{\shortstack{\textbf{MACs}\\(G/s)}} & \multicolumn{8}{c}{\textbf{Evaluation Metrics}} \\
\cmidrule(lr){5-12}
 & & & & \textbf{SDR} & \textbf{SI-SDR} & \textbf{PESQ} & \textbf{STOI} & \textbf{P808} & \textbf{SIG} & \textbf{BAK} & \textbf{OVRL} \\ 
\midrule
\multicolumn{12}{l}{\textit{\textbf{Single-Channel Processing}}} \\
\midrule
1 & No processing & -- & -- & -2.11 & -9.47 & 1.54 & 0.72 & 2.39 & 1.99 & 1.84 & 1.49 \\ 
2 & BSRNN         & 16.9 & 21.15 & 5.31 & -1.38 & 1.94 & 0.79 & 2.62 & 2.61 & 3.29 & 2.18 \\ 
\midrule
\multicolumn{12}{l}{\textit{\textbf{Fixed-Array (Geometry-Specific Upper Bound)}}} \\
\midrule
3 & FaSNet-TAC & 2.7 & 9.76 & 5.13 & -1.03 & 1.69 & 0.72 & 2.46 & 2.36 & 3.14 & 1.95 \\
4 & USES2-comp & 2.5 & 70.26 & 9.36 & 4.78 & 2.56 & 0.87 & 2.77 & 3.14 & 3.49 & 2.65 \\
5 & SpatialNet & 1.2 & 5.99 & 10.06 & 5.23 & 2.56 & 0.87 & 2.75 & 3.11 & 3.58 & 2.67 \\
6 & TF-GridNet & 8.2 & 73.03 & 9.77 & 5.29 & 2.72 & 0.88 & 2.84 & 3.13 & 3.63 & 2.71 \\
\midrule
\multicolumn{12}{l}{\textit{\textbf{Random 4-Mics Array}}} \\
\midrule
7 & FaSNet-TAC & 2.7 & 9.76  & 5.82 & -1.09 & 1.72 & 0.73 & 2.46 & 2.41 & 3.04 & 1.95 \\
8 & USES2-comp & 2.5 & 70.26 & 9.56 & 4.86 & 2.66 & 0.87 & 2.79 & 2.99 & 3.67 & 2.62 \\
9 & SpatialNet & 1.2 & 5.99  & 9.41 & 3.77 & 2.55 & 0.87 & 2.76 & 3.13 & 3.50 & 2.64 \\
10 & TF-GridNet & 8.2 & 73.03 & 9.00 & 3.78 & 2.57 & 0.87 & 2.79 & 3.06 & 3.55 & 2.63 \\
\rowcolor{gray!10} 11 & \textbf{SpatialNet-Geo-DConv (Ours)} & 1.3 & 6.08 & 9.77 & 3.92 & 2.46 & 0.87 & 2.73 & 3.19 & 3.28 & 2.59 \\
\rowcolor{gray!10} 12 & \textbf{TF-GridNet-Geo-DConv (Ours)} & 8.3 & 73.12 & 8.83 & 3.56 & 2.46 & 0.87 & 2.77 & 3.03 & 3.50 & 2.59 \\
\midrule
\multicolumn{12}{l}{\textit{\textbf{Geometry-Invariant (primary comparison)}}} \\
\midrule
13 & FaSNet-TAC & 2.7 & 9.76  & 5.76 & -0.81 & 1.74 & 0.73 & 2.48 & 2.48 & 3.18 & 2.05 \\ 
14 & USES2-comp & 2.5 & 70.26 & 8.62 & \underline{4.17} & \underline{2.52} & \underline{0.86} & 2.76 & 3.02 & 3.50 & 2.56 \\ 
\rowcolor{gray!10} 15 & \textbf{SpatialNet-Geo-DConv (Ours)} & 1.3 & 6.08 & \textbf{9.72} & \textbf{4.22} & 2.48 & \underline{0.86} & \underline{2.77} & \underline{3.16} & \underline{3.51} & \underline{2.68} \\ 
\rowcolor{gray!10} 16 & \textbf{TF-GridNet-Geo-DConv (Ours)} & 8.3 & 73.12 & \underline{9.05} & 3.90 & \textbf{2.59} & \textbf{0.87} & \textbf{2.83} & \textbf{3.21} & \textbf{3.62} & \textbf{2.77} \\ 
\bottomrule
\end{tabular}
\end{table*}

\begin{table}[!t]           
\centering
\caption{Generalization across array topologies (RealMAN) and
         cross-dataset evaluation (CHiME-4). The unprocessed ChiME-4 OVRL score is 1.42.}
\label{tab:generalization_results}
\small
\setlength{\tabcolsep}{4pt}
\renewcommand{\arraystretch}{1.0}

\begin{tabular}{l ccc}
\toprule
\textbf{Array Config} & \textbf{SI-SDR(dB)} & \textbf{PESQ} & \textbf{OVRL} \\

\midrule
\multicolumn{4}{l}{\textbf{USES2-comp}} \\
\quad 1 mic:  \{0\}             & -4.16 & 1.81 & 1.90 \\
\quad 2 mics: \{0,1\}           & 3.55 & 2.46 & 2.54 \\
\quad 5 mics: \{0,1,3,5,7\}     & 4.91 & 2.60 & 2.59 \\
\quad CHiME-4 (cross-dataset)   & -- & -- & 2.55 \\

\midrule
\multicolumn{4}{l}{\textbf{SpatialNet-Geo-DConv (Ours)}} \\
\quad 1 mic:  \{0\}             & 2.30 & 2.15 & 2.57 \\
\quad 2 mics: \{0,1\}           & 3.97 & 2.45 & 2.66 \\
\quad 5 mics: \{0,1,3,5,7\}     & 4.65 & 2.53 & 2.69 \\
\quad CHiME-4 (cross-dataset)   & -- & -- & 2.64 \\

\midrule
\multicolumn{4}{l}{\textbf{TF-GridNet-Geo-DConv (Ours)}} \\
\quad 1 mic:  \{0\}             & 2.76 & 2.37 & 2.70\\
\quad 2 mics: \{0,1\}           & 4.12 & 2.62 & 2.77 \\
\quad 5 mics: \{0,1,3,5,7\}     & 4.54 & 2.67 & 2.79 \\
\quad CHiME-4 (cross-dataset)   & -- & -- & 2.73 \\

\bottomrule
\end{tabular}
\end{table}

Simulated microphone-array datasets often suffer from severe real-world domain mismatch, which leads to performance degradation when models trained on such data are deployed in real acoustic scenarios. To improve model generalization, the Real-recorded and Annotated Microphone Array Speech\&Noise (RealMAN) dataset~\cite{yang_realman_2024} is used to mitigate the simulation-to-real mismatch commonly observed in multichannel speech enhancement and localization. RealMAN provides large-scale real-world recordings captured with a 32-channel high-fidelity microphone array across diverse acoustic environments.

The dataset contains 83.7 hours of multichannel speech (divided into 64.0, 8.1, and 11.6 hours for training, validation, and test) recorded in 32 scenes and 144.5 hours of background noise (divided into 106.3, 16.0 and 22.2 hours for training, validation and test) recorded in 31 scenes. Recording environments include indoor, outdoor, semi-outdoor, and transportation scenarios. Sub-arrays extracted from the 32-channel array enable training and evaluation under variable array configurations. 

Recording environments include indoor, outdoor, semi-outdoor, and transportation scenarios. Speaker locations are annotated using an omnidirectional fisheye camera that automatically tracks the loudspeaker. For SE, the direct-path signal—obtained by filtering the source speech with an estimated direct-path propagation filter—is used as the target clean speech. 
\subsection{Implementation Setup}

All models are trained on the same RealMAN dataset. For computational efficiency, all recordings are resampled to 8 kHz. A 256-point window and 128-sample frame shift are used to compute the STFT. During both training and evaluation, the input signals are segmented into 4-second utterances. To ensure a fair comparison, the segmentation boundaries are kept strictly identical across all models during testing.

BSRNN~\cite{luo2023bsrnn} is selected as the representative single-channel speech enhancement (SE) baseline. For array-agnostic models, FaSNet-TAC~\cite{luo_end--end_2020} and USES2-comp~\cite{zhang_improving_2024} are adopted for comparison. In addition, SpatialNet~\cite{Quan2024SpatialNet} and TF-GridNet~\cite{wang2023tf-gridnet} are included as representative fixed-array SE baselines.

For the proposed Geo-DConv architecture, the basis dimension $b$ is set to 8, the number of output channels $O$ is 16, and spherical coordinates are adopted. The number of frequency bands in PE is configured as 6. For the TACT block, $d_{\text{hidden}}$ is set to 64, and a Transformer layer with 4 heads and 2 layers is employed. SDR, SI-SDR~\cite{jepsen2025sisdr}, PESQ~\cite{Rix_2001_PESQ}, STOI~\cite{Taal_2010_STOI}, and DNSMOS ~\cite{reddy2022dnsmos} are employed as evaluation metrics in the subsequent experiments.

\section{Results and Analysis}
\subsection{Impact of Fixed vs. Random Array Training}
Two data-feeding strategies are used to evaluate the impact of array geometry configuration during training: utilizing a random 4-mic array versus a fixed-geometry 4-mic array. During the testing phase, all models are evaluated on the same fixed-geometry 4-mic array (microphone indices [0, 1, 5, 9]). As demonstrated in Table~\ref{tab:main_results} (Nos. 3–10), fixed-array methods such as SpatialNet and TF-GridNet show the upper bound after training with the specific array. However, these models lead to a significant degradation in enhancement performance, especially in signal-level metrics like SDR and SI-SDR, when trained with random arrays, as they fail to reliably leverage the array structure.

For array-agnostic models like FaSNet-TAC and USES2-comp, random arrays occasionally yield slightly superior results. Because the structural design of these models predominantly focuses on single-channel spectra, they are inherently insensitive to variations in array geometry. Consequently, training with random arrays acts as a form of data regularization, subtly enhancing overall performance. In contrast, for fixed-array algorithms, training with a fixed-geometry array allows the model to implicitly capture the geometric information embedded within the data, thereby enabling more effective spatial modeling.

\subsection{Performance Comparison of Array-Invariant Methods}
Previous studies have confirmed that fixed-array algorithms yield superior enhancement for specific configurations, as geometric information is crucial for their performance. By incorporating the proposed Geo-DConv structure, the fixed-array algorithm can be adapted to handle randomized array configurations. As shown in Table~\ref{tab:main_results}, the proposed SpatialNet-Geo-DConv and TF-GridNet-Geo-DConv outperform the previous FaSNet-TAC previous FaSNet-TAC and USES2-comp models. In particular, SpatialNet-Geo-DConv achieves performance comparable to USES2-comp, but with significantly lower computational cost and fewer parameters. Through the comparison between Nos.11–12 and Nos.15–16, performance is further bolstered by training with a randomized number of microphones. This demonstrates that variable input channel counts help the model better learn to generate dynamic convolution weights. 

To evaluate the adaptability of different Geometry-Invariant methods to various array structures, experiments are performed on varying numbers of microphones and array geometries, as shown in Table~\ref{tab:generalization_results}.When only a single microphone is used, all methods reduce to single-channel SE.The proposed SpatialNet-Geo-DConv and TF-GridNet-Geo-DConv achieve significantly better enhancement performance than USES2-comp. As the number of microphones increases, all methods exhibit consistent performance improvements.Notably, the Geometry-Invariant methods are trained with a maximum of 4 microphones but can generalize well to 5-microphone and 6-microphone setups, including the 6-microphone real-recorded CHiME-4 test set~\cite{Vincent2017chime4}. The results demonstrate excellent generalization ability, with the DNSMOS OVRL improved from 1.42 to 2.64 and 2.73 for the two proposed methods on CHiME-4, respectively. 

Although the models are trained only on the RealMAN dataset with a maximum of 4 microphones,
it can generalize well to the 6-microphone real-world setup in CHiME-4 without any fine-tuning.
This confirms that the proposed architecture can effectively capture general geometry-aware spatial patterns
rather than overfitting to a specific array structure.
Moreover, training on real-recorded data alleviates the sim-to-real domain mismatch,
which is critical for practical deployment in real acoustic environments.

\section{Conclusions}
This paper reveals that while traditional fixed-array algorithms are inherently constrained by specific geometries, they can more effectively exploit spatial information. Building upon these insights, we propose the Geometry-Aware Dynamic Convolution module, which not only addresses the limitation of conventional convolutional layers for variable input dimensions but also explicitly incorporates array geometric priors. By introducing this module, conventional fixed-array methods can be converted into array-invariant SE systems, achieving performance improvements over existing array-invariant approaches. This work provides a new perspective for achieving array-invariant SE and represents an initial attempt to leverage explicit array structure information to assist multi-channel SE.

\section{Acknowledgments}
This work was supported in part by China NSFC project under Grants No. U25A20409, and in part by SJTU Med-X (Medicine \& Engineering) Translational Research Grant (YG2025LC09).

\section{Generative AI Use Disclosure}
Generative AI tools were used solely for language polishing and grammatical improvement in the writing process. All scientific content, ideas, analysis, and conclusions are original and fully authored by the researchers. The authors take full responsibility for the final manuscript.

\bibliographystyle{IEEEtran}
\bibliography{mybib}

\end{document}